\documentclass[notitlepage,aps,pre,twocolumn,a4paper,twoside]{revtex4-1}
\usepackage{amsmath,amssymb,amsthm,mathrsfs,amsfonts,dsfont}
\usepackage[utf8]{inputenc}

\def\be{\begin{equation}}
\def\ee{\end{equation}}

\long\def\symbolfootnote[#1]#2{\begingroup%
\def\thefootnote{\fnsymbol{footnote}}\footnote[#1]{#2}\endgroup}

\newlength{\defbaselineskip}
\newcommand{\setlinespacing}[1]%
           {\setlength{\baselineskip}{#1 \defbaselineskip}}

\newcommand{\singlespacing}{\setlength{\baselineskip}{\defbaselineskip}}

\theoremstyle{plain}
\newtheorem{thm}{Theorem}[section]

\theoremstyle{definition}
\newtheorem{defn}[thm]{Definition}

\theoremstyle{remark}

\begin{document}

\title{Quantum theory as a critical regime of language dynamics}
\author{Alexei Grinbaum}\email{alexei.grinbaum@cea.fr} \affiliation{CEA-Saclay/IRFU/LARSIM, 91191 Gif-sur-Yvette, France}
\begin{abstract}\noindent
Some mathematical theories in physics justify their explanatory superiority over earlier formalisms by the clarity of their postulates. In particular, axiomatic reconstructions drive home the importance of the composition rule and the continuity assumption as two pillars of quantum theory. Our approach sits on these pillars and combines new mathematics with a testable prediction. If the observer is defined by a limit on string complexity, information dynamics leads to an emergent continuous model in the critical regime. Restricting it to a family of binary codes describing `bipartite systems,' we find strong evidence of an upper bound on bipartite correlations equal to $2.82537$. This is measurably different from the Tsirelson bound. The Hilbert space formalism emerges from this mathematical investigation as an effective description of a fundamental discrete theory in the critical regime.
\end{abstract}
\maketitle

\section{Mathematics guides understanding}\label{guidesect}

The goal of mathematical improvements of physical theories is to achieve explanatory superiority over previous formalisms. This does not always come with a more practical method of doing calculations. 
In this article we will suggest a novel mathematical approach to the problem of quantum mechanical observer. Before doing so, we make a methodological point on a historic example: while von Neumann's quantum logic and Weyl's theory of relativity stood at a pragmatic disadvantage with respect to earlier formalisms, they maintained the upper hand in explaining physics.

A key episode in the history of quantum theory --- Werner Heisenberg's introduction of matrix mechanics --- happened due to Max Born's knowledge of matrix multiplication. Born told Heisenberg about a mathematical formalism yet unused in physics, which corresponded precisely to Heisenberg's rules for calculating atomic spectra~\cite{unreas}. This was but a first instance of bringing new mathematics into quantum theory. A little later John von Neumann proposed a different mathematical foundation based on the theory of operators in the Hilbert space~\cite{HvNN,vN32}, whose success was due to the capacity of operator theory to generalize the rules of matrix multiplication. The pragmatism and simplicity of Heisenberg's initial rules were manifestly not sufficient to install them as fundamental ingredients of quantum theory. Something else was needed: a satisfactory explanation. 
 
Constant search for a deeper mathematical foundation led von Neumann to doubt even the Hilbert space formalism. In 1935, he wrote to Garrett Birkhoff: ``I would like to make a confession which may seem immoral. I do not believe absolutely in Hilbert space any more''~\cite[p.~59]{vNletters}. Together with Birkhoff, von Neumann initiated a program of quantum logic, replacing the Hilbert space axioms by a discrete structure later called an orthomodular lattice~\cite{bvn}. The reception of von Neumann's work, even among theoretical physicists, was rather cool~\cite{GorelikBron}. Many thought that it was too far removed from the possibility of doing calculations; quantum logic even more so than operator theory. Yet von Neumann's program has made a lasting contribution to quantum theory by providing its first set of fundamental axioms. Although his and later reconstructions did not produce new predictions, the hope has been vindicated that the axiomatic approach would provide a clearer understanding of the existing theory~\cite{grinbjps}.

A somewhat similar situation occurred in spacetime physics. Einstein's general theory of relativity, a textbook example of how mathematical methods can be successfully applied to physics, was based on the metric tensor. In opposition to Einstein, Hermann Weyl pursued an approach based on the affine connection~\cite{Weyl}. His was a version of Riemannian geometry with a different main concept and, while Weyl was gradually surrendering all claims in favor of his theory as a practical replacement of Einstein's, he still ``marshaled a number of aesthetic and `philosophical' arguments that he believed recommended his theory over that of Einstein''~\cite[p.~160]{Ryckman}. Perhaps the expectation of a pragmatic advantage was vain, but conceptual superiority of his theory over Einstein's was, according to Weyl, beyond doubt. Why, then, did his approach stumble? We believe that Weyl's formalism was too close to Einstein's: it used the same fundamental mathematics of Riemannian geometry, even if the metric tensor was replaced with a connection. Most physicists only saw a minor improvement in Weyl's idea. Whenever mathematics purports a new explanation of physical theory, it should, in our view, rely on a framework profoundly different from the common one.

Unlike Weyl's theory, quantum logic used a radically new formalism. Mackey~\cite{mackey57} and Piron~\cite{piron}, among others, gave examples of mathematical derivations of the Hilbert space formalism from an orthomodular lattice with additional assumptions. Such reconstructions provide an important insight: an assumption of continuity is a necessary, but not a sufficient, ingredient of quantum mechanical axiomatic systems~\cite{zieler,land,Masanes,Chiri}. Differently worded continuity assumptions exist in every reconstruction: a prominent representative is the existence of a continuous reversible transformation between any two pure states of the system~\cite{BruknerDakic}. On their own, however, these assumption are insufficient for reconstructing quantum theory, as demonstrated by $C^{\ast}$-algebraic approaches~\cite{Bub}. 

Quantum theory has emerged from the reconstruction program, not only as a description of individual systems with continuous state spaces, but also requiring an extra axiom about how such systems compose~\cite{popescu,hardy}. This second insight must be complemented with a quantitative bound on the amount of correlations given by the Bell inequalities~\cite{Bell1} and explored in postquantum models~\cite{spekkens2,barrett2,popescu2014}. There exists a fundamental fact about nature: the amount of correlations between distant subsystems is limited by a non-classical bound, e.g., the Tsirelson bound for bipartite correlations~\cite{Tsirel}. In our view, all mathematical alternatives to the Hilbert space formalism must strive to predict its empirical value.

Another avenue leading to the question about the mathematics of quantum theory begins with the problem of observer. Quantum theory says nothing about its physical composition. It only describes the observer's information, which must be somehow registered. Hugh Everett argued that observers are characterized by their memory, i.e., ``parts... whose states are in correspondence with past experience''~\cite{everett}. It seems reasonable to assume that differently constituted observers with the same memory size will have equal capacity to register measurement results. This is because quantum mechanics uses abstract mathematics: it deals with \textit{observers} possessing \textit{information} about \textit{systems}, assuming nothing about material counterparts of these notions. Can this level of abstraction be used to describe observers? We find a somewhat puzzling answer to this question in the work of Niels Bohr. Over time Bohr ``became more and more convinced of the need of a symbolization if one wants to express the latest results of physics''~\cite{jammer}. He had already written extensively on the problem of objective description, but he only connected it in 1958 with the choice of mathematical formalism for quantum theory: ``The use of mathematical symbols secures the unambiguity of definition required for objective description''~\cite{bohr1958}. What exactly Bohr meant remains unclear. It is unlikely that his point was that quantum theory must rely on the Hilbert space, by then a standard tool. Had it been so, Bohr could have named the Hilbert space explicitly, yet he only vaguely referred to ``mathematical symbols''. In the same text Bohr also rejected the Schr\"odinger wave function as a candidate mathematical tool. It is conceivable that Bohr's view was that such mathematical symbols remained to be found. If so, their discovery would purportedly guarantee the unambiguity of communication and secure the objectivity of description. It would then account to a ``common-language'' basis of physical theory, in line with Bohr's well-known insistence on the role of classical concepts~\cite{bohr1934}.

Our attempt to clarify the meaning of Bohr's statement leads to two questions. First, is there a mathematical framework that includes both ambiguous and unambiguous descriptions? In Section~\ref{sect_codes} we introduce such a framework based on the algebraic coding theory. This theory provides a general model of communication and deals mathematically with errors or ambiguity. Second, how is quantum theory different from all other unambiguous descriptions? As emphasized above, it must obey a condition of continuity and a bound on bipartite correlations.

The use of coding theory is enabled by the definition of observer in information-theoretic terms introduced in Section~\ref{sect_complex}. It involves a limit on the complexity of strings, which (to use a common-language expression) the observer can `store and handle'. Strings contain all descriptions of states allowed by quantum theory but also much more: they may not refer to systems or be interpretable in terms of preparations or measurements. Using the work of Manin and Marcolli, we show that symbolic dynamics on such strings leads to an emergent continuous model in the critical regime (Section~\ref{sect_cont}). Restricting this model to a subfamily of `quantum' binary codes describing `bipartite systems' (Section~\ref{sect_corr}), we find strong evidence of an upper bound on bipartite correlations equal to $2.82537$. The difference between this number and the Tsirelson bound $2\sqrt{2}$ can be tested. The Hilbert space formalism, then, emerges from this mathematical approach as an effective description of a fundamental discrete theory of `quantum' languages in the critical regime, somewhat similarly to the description of phase transitions by the effective Landau theory.

\section{Codes}\label{sect_codes}

Communication is based on encoding messages that are transmitted in suitable codes using an alphabet shared between communicating parties. An alphabet is a finite set $A$ of cardinality $q\geq 2$. A code is a subset $C \subset A^n$ consisting of some of the words of length $n\geq 1$. A language is an ensemble of codes of different lengths using the same alphabet.
As an example, take the alphabet $A=F_q$, the finite field of $q$ elements. Linear subspaces of $F_q^n$ give rise to codes called linear. Linearity provides such codes with extra structure. Another example is given by binary codes of length $n$ based on a two-letter alphabet, say, $\{0,1\}$. Strings of zeros and ones of arbitrary length belong to a language formed by binary codes with different values of $n$.

In full generality, nothing can be stipulated about message semantics, material support of the encoding and decoding operations or their practical efficiency. One can observe, however, that decoding a message is less prone to error if the number of words in the code is small. On the other hand, reducing the number of code words requires the words to be longer. The number \begin{equation}R=\frac{\log _q \#C}{n}\end{equation}is called the (transmission) rate of code $C$.

One can associate a fractal to any code in the following way~\cite{Manin1,Manin2}. Define a rarified interval $\left(0,1\right) _q = \left[ 0,1 \right] \setminus \{ m/{q^n} | m,n \in \mathds{Z}\}$. Points $x=(x_1,\ldots , x_n)\in \left( 0,1\right) ^n _q$ can be identified with $(\infty \times n)$ matrices whose $k$-th column is the $q$-ary decomposition of $x_k$. For a code $C$ define $S_C \subset \left( 0,1\right) ^n _q$ as the set of all matrices with rows in $C$. It is a Sierpinski fractal and its Hausdorff dimension is $R$. The closure of $S_C$ inside the cube $[0,1]^n$ includes the rational points with $q$-ary digits. This new fractal $\hat{S}_C$ is a metric space in the induced topology from $[0,1]^n.$ Now consider a family of codes $C_r$ of $\# C_r = q^{k_r}$ words of length $n_r$, with rates:
\begin{equation}
\frac{k_r}{n_r}\nearrow R.
\end{equation}
They define a fractal $S_R=\bigcup _r S_{C_r}$ of Hausdorff dimension $\mathrm{dim} _H (S_R) = R$.

\section{Bounded complexity}\label{sect_complex}

Any observer's memory is limited in size. While their material constitution may be radically different, different observers with the same memory size should demonstrate similar performance in handling information. This intuition serves as a motivation for the following information-theoretic definition of observer.
\begin{defn}\label{defobs} An observer is a subset of strings of bounded complexity, i.e., strings compressible below a certain threshold. \end{defn}
This limit can be viewed as the length of observer's memory. If a string has high complexity, it cannot describe an observer with a memory smaller that the minimal length required to store it; but it remains admissible for an observer with a larger memory.

Definition~\ref{defobs} requires a notion of string complexity independent of the observer's material organization. Kolmogorov complexity is a suitable candidate. It has already been used in fundamental physics, e.g. by Zurek who argued that physical entropy should be defined as a sum of Shannon entropy $H$ and algorithmic randomness of available information~\cite{Zurek1,Zurek2,Zurek3}. The latter was to be understood as information contained in a `binary image' of the state of the system, defined as Kolmogorov complexity $K$ of the shortest program able to generate it. When the state of the system is known sufficiently well, $K$ supplies the main contribution to entropy. Zurek argued that this algorithmic component of physical entropy can be made observer-independent by discretizing the system on a family of grids that are concisely describable by a universal computer. To extend Zurek's understanding of an observer who `interprets' a string as containing information about the state of the system, we take such strings to be information-theoretic primitives. In our approach, strings do not necessarily have an interpretation as states of a system: they define the observer. Mathematical analysis proceeds, however, independently of the choice of Zurek's or our interpretation. For a set of strings that are code words of code $C$ with rate $R$, the lower Kolmogorov complexity satisfies~\cite{Manin2}:
\begin{equation}
\sup _{x\in \hat{S}_C} \kappa (x) = R.
\end{equation}
For all words $x\in S_{R}$ in a language formed by codes $C_r$, the lower Kolmogorov complexity is bounded by $\kappa (x) \leq R$.
Hence the closure $\hat{S}_R$ of the fractal $S_R$ is a metric space that describes the handling of words of bounded Kolmogorov complexity. It is a `minimal' geometric structure corresponding to the notion of observer.

\section{Critical language dynamics}\label{sect_cont}

A change in observer's information can be modeled via dynamical evolution on the fractal set $S_R$. In quantum theory, new information enters when a projective or a POVM measurement produce a new string in observer's memory. Taking inspiration from Manin and Marcolli~\cite{Manin2}, we represent this process as a statistical mechanical system evolving on the set of all possible strings in codes $C_r$. A change in observer's information corresponds to a change in the `occupation numbers' $\lambda_a$ of words $a\in \bigcup _r C_r$. The evolution of $\lambda_a$ is described via Hamiltonian dynamics on the Fock space:
\begin{equation}
H_{stat} \epsilon _{a_1 \ldots a_m} = (\lambda _{a_1}+\ldots +\lambda _{a_m})\epsilon _{a_1 \ldots a_m},
\label{HamilManin}
\end{equation}
with the Keane `ergodicity' condition:
\begin{equation}
\sum \limits_{a\in \cup _r C_r} e^{-R\lambda _a} =1,
\end{equation}
where vectors $\epsilon _{a_1 \ldots a_m}$ belong to the Fock space representation $l^{2} (W(\bigcup _r C_r))$ of the set $W$ of all words in the codes $C_r$. To be precise, the Fock space is a representation of the algebra defined below in (\ref{projectdef}); at this stage it is justified by the completeness of $W$, which by construction includes all possible observer information states. If observer's information remains within $W$, then the Keane condition gives a meaning to the weights $\lambda_a$ as normalized logarithms of inverse probabilities that $a$ is stored in observer's memory. This evolution has a partition function:
\begin{equation}
Z(\beta)=\frac{1}{1-\sum \limits_{a\in \cup _r C_r}e^{-\beta\lambda _a}}.
\end{equation}

Manin and Marcolli show that at the critical temperature (equivalently, string complexity) $\beta = R$, the behaviour of this system is given by a KMS state on an algebra respecting unitarity~\cite{Manin2}. This algebra is built out of the geometric object, namely the fractal $\hat{S} _C$, as follows. Consider characteristic functions $\chi _{\hat{S} _C (w)}$, where $w=w_1\ldots w_m$ runs over finite words composed of $w_i\in C$ and $\hat{S} _C (w)$ denotes the subset of infinite words $x\in \hat{S}_C$ that begin with $w$. These functions can be identified with the range projections \begin{equation}P_w=T_w T^*_w=T_{w_1}\ldots T_{w_m}T^*_{w_m}\ldots T^*_{w_1}.\label{projectdef}\end{equation}
At the low temperature $\beta > R$ there exists a unique type $I_\infty$ KMS-state $\phi _R$ on the statistical system of codes, which is a Toeplitz-Cuntz algebra with time evolution: \begin{equation}\sigma _t (T_w)=q^{itn}T_w.\end{equation} The partition function is: \begin{equation}Z_C(\beta)=(1-q^{(R-\beta)n})^{-1}.\end{equation} However the isometries in the algebra do not add up to unity. Only at the critical temperature $\beta=R$, where a phase transition occurs for all codes $C_r$, is there a unique KMS state on the Cuntz algebra, i.e., an algebra such that, importantly for our argument, isometries add up to unity: $\sum _{a} T_a T_a ^* =1.$

Critical behavior of the original discrete linguistic model is described at $\beta=R$ by a field theory on the metric space $\hat{S}_R$, which obeys unitarity. By construction, this fractal also has scaling symmetry. This yields a field theory respecting scaling and unitarity. While there has been some discussion of models that are scale invariant but not conformal, we assume that, in agreement with Polyakov's general conjecture~\cite{Polyakov}, this field theory is conformal. The field it describes is clearly an emergent phenomenon, for its underlying dynamics is given in terms of codes. However, within the conformal field theory this field, now a basic object, is to be considered fundamental. Due to the properties of continuity, unitarity and to the geometric character of its state space, the conformal model becomes a tentative candidate for a reconstruction of quantum theory.

\section{Amount of correlations}\label{sect_corr}

Since quaternionic quantum mechanics or, in some limited cases, real-number quantum mechanics can be represented in the Hilbert space~\cite{adler}, one should expect that continuity and unitarity alone do not single out quantum theory. In other words, the conformal model of Section~\ref{sect_cont} likely contains more than a description of `quantum' languages. In this section, we do not seek to provide a necessary and sufficient condition that
selects only code words generated by quantum theory. Rather, we pick out a particular example, namely a class of models corresponding to the critical regime of binary codes describing measurements on bipartite quantum systems in the usual 3-dimensional Euclidean space.

First we define an informational analog of `bipartite.' In quantum theory, subsystems that are entangled can be materially different, but they are described by the same number of entangled degrees of freedom. Their informational content is represented by strings of identical complexity. For example, measurements in a CHSH-type experiment produce binary strings of results for a choice of $\sigma_x,\sigma_y,\sigma_z$ measurements. The no-signalling condition implies that the probability of 0 on Alice's side is independent of Bob's settings, and vice versa. Hence the strings resemble Bernoulli distributions with a Kolmogorov complexity equal to the binary entropy of the probability of 0, plus a correction due to the existence of non-zero mutual information between Alice's and Bob's outputs. Since both sides enter symmetrically in the CHSH inequality, this correction to Kolmogorov complexity is \textit{a priori} the same on Alice's and Bob's sides. We use this argument to replace Eq.~(\ref{HamilManin}) with a class of Hamiltonians assumed to describe a `bipartite system' in the framework of codes.

The Kolmogorov order is an arrangement of words $a_i\in \bigcup _r C_r$ in the increasing order of complexity~\cite{Manin3}. It is not computable and it differs radically from any numbering of $a_i$ based on the Hamming distance in the codes $C_r$. Words that are adjacent in the Kolmogorov order have the same complexity. We now select an Ising-type Hamiltonian:
\begin{equation}\label{HamilKolm}
H_{2}=-\sum_{ij} a_i \times a_j,
\end{equation}
as a dynamical model on the language that describes bipartite quantum systems. The sum is taken over $N$ neighbors in the Kolmogorov order, i.e. all strings of identical complexity. The result of multiplication on binary words is a new word with letters isomorphic to multiplication results in a two-element group $\{\pm 1\}$. Hence, for a two-letter alphabet $\{a,b\}$,
\begin{equation}
a\times a = b \times b = b,\quad a \times b = b \times a = a.
\end{equation}
A binary language with $N=6$ using $H_2$ gives rises to information dynamics which is, on the one hand, equivalent to information dynamics of a bipartite quantum system and, on the other hand, equivalent to the dynamics of a 3-dim Ising model. This is because a class of Hamiltonians with $N=6$ has the same number of terms as in three spatial dimensions, although the codes that belong to this class are uncomputable due to the properties of Kolmogorov complexity. Plainly, one cannot tell which binary codes give rise to the $N=6$ situation nor should one expect that Hamiltonians $H_{stat}$ and $H_2$ belong to the same universality class. However, the equivalence of (\ref{HamilKolm}) with a 3-dim Ising model suggests that, just like the Ising model itself, the Hamiltonian $H_2$ also exhibits critical behaviour described by a conformal field theory.

As it is usually the case in statistical mechanics, critical regime can be studied without knowing the details of the dynamics. Correlations of order $2$ in this regime are described by the lowest-dimensional even primary scalar $\epsilon = \sigma \times \sigma$ in the conformal field theory. This field is symmetric; hence it presents a good candidate to describe the symmetry of bipartite correlations in the CHSH inequality under the switch between Alice and Bob. Following the above intuition, we assume that it provides a description of `bipartiteness' within the conformal model. The operator dimension of $\epsilon$ is \begin{equation}\Delta _\epsilon = 3 - \frac{1}{\nu},\end{equation} where $\nu$ is a well-known critical exponent describing the correlation length~\cite{Rychkov2012}.

The 3-dim Ising equivalence has its limitations since the true metric space of code evolution is not flat space but the fractal $\hat{S}_C$. Still, it provides significant evidence that $H_2$ has a critical regime. Further, exponential character of the mapping that links the fractal embedded in the unit cube with flat Euclidean space hints at the existence of a connection between the critical behaviours of the Ising model and the code. The correlation length in the fractal representation of a language describes a logarithmic distance from which words are brought in clusters of equal complexity by the Kolmogorov reordering. If $H_2$ exhibits a critical behaviour similar to that of $H_{stat}$, then correlations in the critical regime at string complexity $\beta=R$ come from the entire fractal. The Ising analogy with the scaling of the correlator of the lowest primary even field suggests a power law for the amount of correlations on the words of equal complexity:
\begin{equation} \label{scaletrans}
\langle \epsilon (a) \epsilon (0) \rangle \sim a^{-2\Delta_\epsilon}.
\end{equation}
We conjecture that, due to the exponential mapping between spaces, the corresponding correlations in the fractal are limited by the logarithm of the RHS of (\ref{scaletrans}). Their maximum strength $2\Delta_\epsilon$ can be computed based on the value $\nu=0.62999(5)$ in~\cite{Rychkov2014}:
\begin{equation}
2\Delta_\epsilon=2.82537(2).
\end{equation}

\section{Conclusion}\label{conclusion}

Historically, quantum logical reconstructions of quantum theory drive home the importance of the assumptions of continuity and composition rule. These are two pillars of quantum theory. Freely interpreting Bohr's dictum that the unambiguous communication of measurement results requires a mathematical formulation, we proposed a mathematical framework that sits on these pillars and the idea that the observer is defined by the limited string complexity.
The result is a conjecture on the amount of bipartite correlations slightly different from the Tsirelson bound, but consistent with available experimental results $S=\Delta_\epsilon +2\simeq 3.41267 \leq 3.426\pm 0.016$~\cite{nawareg} and $2\Delta_\epsilon \simeq 2.82534 \leq 2.827 \pm 0.017$~\cite{kwiat}.

Our argument crucially relies on the assumption that the number of strings possessing the same complexity after uncomputable Kolmogorov reordering is $N=6$. This does not need to be so for all codes. Codes with $N=4$ correspond to 2-dim Ising interaction ($\nu=1$) and give rise to the classical bound on bipartite correlations $2\Delta_\epsilon=2\cdot 1=2$. Codes with $N=8$ correspond to 4-dim Ising interaction ($\nu=1/2$) and give rise to the Popescu-Rorlich maximum correlation $2\Delta_\epsilon=2\cdot (4-\frac{1}{1/2})=4$. It is not clear whether binary codes endowed with critical dynamics exist for other values of $N$ and, if they do, what meaning they may have.

Although our model is highly speculative, we believe that it demonstrates the interest to explore quantum theory via novel mathematical formalisms. As Wittgenstein said, ``A particular method of symbolizing may be unimportant, but it is always important that this is a possible method of symbolizing. [This] possibility\ldots reveals something about the nature of the world''~\cite{Witt}.

\singlespacing\footnotesize

%

\end{document}